\documentclass[12pt,a4paper]{article}
\usepackage{a4wide}
\usepackage{amsmath}
\usepackage{amsfonts}
\usepackage{cite}
\usepackage{graphicx}
\begin{document}

\title{On the separability criterion of bipartite states with certain non-Hermitian operators}
\author{{N. Ananth$^a$, V. K. Chandrasekar$^b$ and  M. Senthilvelan$^a$ }\\
{\small $^a$Centre for Nonlinear Dynamics, School of Physics, Bharathidasan University,} \\{\small Tiruchirappalli - 620024, Tamilnadu, India} 
\\{\small $^b$Centre for Nonlinear Science \& Engineering, School of Electrical \& Electronics Engineering,} \\
{\small SASTRA University, Thanjavur - 613401, Tamilnadu, India}}
\date{}
\maketitle

\begin{abstract}
We construct a density matrix whose elements are written in terms of 
expectation values of non-Hermitian operators and their products 
for arbitrary dimensional bipartite states. 
We then show that any expression which involves matrix elements can be reformulated by the expectation 
values of these non-Hermitian operators and vice versa. 
We consider the condition of pure states and pure product states and rewrite them in terms of expectation values 
and density matrix elements respectively. 
We utilize expectation values of these operators to present the condition for separability 
of ${C}^d \otimes {C}^d$ bipartite states. With the help of our separability criterion we detect entanglement 
in certain classes of higher dimensional bipartite states.
\end{abstract}

\section{Introduction}
\label{intro}
During the past two decades considerable progress has been made to characterize 
the pure and mixed states and their entanglement with multifaceted applications 
in the field of quantum computation and information \cite{niel2000,peres2002}. One of the fundamental tasks 
in quantum information theory is to detect entanglement in quantum states. In general, an entanglement can be 
identified from the separability criterion. Several necessary and/or sufficient separability 
criteria have been proposed in the literature in order to capture the entanglement 
of bipartite states \cite{horo2009,guhne2009,dohe2004,brub2002}. Among these criteria,  
the positive partial transpose (PPT) plays a vital role in detecting entanglement in   
$2\otimes2$ and $2\otimes3$ bipartite states  \cite{peres1996,horo1996}. 
The separability criterion in terms of the range of density matrix and the computable 
cross-norm criterion (CCNR) work efficiently for the $3\otimes 3$ and $2\otimes 4$ mixed states 
whereas PPT fails to detect entanglement in these states \cite{horo1997,rudo2005}.
Various approaches to construct Bell inequalities were also proposed 
for pure two qudit systems \cite{gisin1992} and higher dimensional bipartite systems \cite{collins2002,li2010}.  
The diagonalization criterion and Bell-type inequalities are put forward for the separability of $M\otimes N$ 
and $2\otimes d$ bipartite states in Refs.\cite{ha2010,zhao2011}. 
A class of $d\otimes d$ bipartite PPT states were proposed in \cite{chrus2006} in 
connection with the indecomposability of positive maps. However, a strong separability criteria is yet to be proposed for 
arbitrary dimensional bipartite states. 

In the study of entanglement of bipartite states, a few works have been devoted to analyze 
the entanglement with the help of non-Hermitian operators. For example, 
T\'oth et al. have derived an inequality to detect entanglement in two-mode continuous systems using  
number operator and mode annihilation operator \cite{toth2003}.
Shchukin and Vogel have derived general entanglement conditions 
for continuous bipartite states \cite{shch2005}. Subsequently, Hillery and Zubairy have developed certain entanglement conditions for 
two-mode systems by considering mode creation and annihilation operators \cite{hill2006}. 
In a later work \cite{hill2006a} the same authors were given a wide range of applications of the entanglement 
conditions \cite{hill2006}. 
Interestingly, these conditions were further strengthened to detect entanglement not just 
between field modes but also between atom and field modes or between groups of atoms, 
see for example Ref.\cite{hill2009}. In addition to the above, non-Hermitian operators were also employed to demonstrate the 
entanglement in multipartite states \cite{hill2010}. The above studies reveal that non-Hermitian operators can also be utilized  
to characterize the quantum states in a new way. 

In this paper, we detect entanglement with the help of non-Hermitian operators. 
The aim of this work is to derive a separability criterion in terms of expectation values of non-Hermitian operators 
and their products 
for the higher dimensional bipartite states. 
To achieve this goal, we construct a qudit density matrix, whose elements are replaced by the 
expectation values of non-Hermitian operators and their products. We then derive the same form of matrix for bipartite states 
by implementing a tensor product between expectation value matrices of two qudit states.
Our analysis shows that higher dimensional states can easily be represented 
in terms of non-Hermitian operators. 
We then show that any expression which involves matrix elements can be reformulated by the expectation 
values of these non-Hermitian operators and vice versa. 
To demonstrate this, we rewrite the condition of pure states, Tr$(\rho^2)$$=1$, in terms 
of expectation values of non-Hermitian operators and their products. 
We then consider the condition of pure product state and reformulate it in terms of density matrix elements 
from the expectation value of non-Hermitian operators.
We also present an operational form of partial transposition operation.
Proceeding further, we formulate a separability condition to the mixed bipartite states in terms of density matrix elements 
using the fact that they can be rewritten in terms of expectation values. To derive separability criterion for higher dimensional bipartite states,  
we consider $\langle A^{\dagger} B^{\dagger} \rangle$ and $\langle A^{\dagger} B \rangle$, where  
the non-Hermitian operators $A$ and $B$ act on first and second subsystem respectively,  
from which we find the necessary elements which are suitable for the Werner-like states and Isotropic like states \cite{chrus2006}.  
We then use the relation of these elements with diagonal elements in the density matrix \cite{guhne2010,gao2011} 
and obtain a condition for entanglement of ${C}^d \otimes {C}^d$ bipartite states in the form of an inequality. 
We also demonstrate the validity of our formulation by considering four different states.    
 
We organize our work as follows. In the following section, we construct a density matrix for $M\otimes N$ 
bipartite states whose elements are in the form of expectation values of certain non-Hermitian operators and 
their products. In Sect.\ref{sec3} we show the utilization of expectation value matrix for few simple cases. 
We then derive the separability condition for ${C}^d \otimes {C}^d$ bipartite states in Sect.\ref{sec4} and 
demonstrate the applicability of our method by considering four different states in Sect.\ref{sec5}. 
Finally, we summarize the conclusion in Sect.\ref{sec6}. 

\section{Density matrix in terms of non-Hermitian operators}
\label{sec2}
Let $\rho^1$ and $\rho^2$ denote the states of two subsystems on the Hilbert space $\mathcal{H}^1$ 
and $\mathcal{H}^2$ respectively. The state of the composite system is then 
$\rho \in \mathcal{H}^1 \otimes \mathcal{H}^2$. 
It is known that any separable state can be expressed in the form \cite{lewn2000}.
\begin{subequations}
\begin{align}
\label{0} \rho =& \rho^{1} ~\otimes~\rho^{2}, \\
\label{1} \rho =& \sum_{i} p_i~\rho_{i}^{1} ~\otimes~\rho_{i}^{2},
\end{align} 
\end{subequations}
where $p_i > 0$ and $\sum_{i} p_i = 1$, for pure and mixed state respectively.
The density matrix representation of a qudit state is given by 
\begin{align}
\label{3a} \rho^k= 
\begin{pmatrix}
\big\langle 0|\rho^k |0 \big\rangle & \big\langle 0| \rho^k |1 \big\rangle & \big\langle 0| \rho^k |2 \big\rangle & \ldots  & \big\langle 0| \rho^k |m \big\rangle \\
\big\langle 1|\rho^k |0 \big\rangle & \big\langle 1| \rho^k |1 \big\rangle & \big\langle 1| \rho^k |2 \big\rangle & \ldots  & \big\langle 1| \rho^k |m \big\rangle \\
\big\langle 2|\rho^k |0 \big\rangle & \big\langle 2| \rho^k |1 \big\rangle & \big\langle 2| \rho^k |2 \big\rangle & \ldots  & \big\langle 2| \rho^k |m \big\rangle \\
\vdots & \vdots  & \vdots & \ddots & \vdots \\
\big\langle m|\rho^k |0 \big\rangle & \big\langle m| \rho^k |1 \big\rangle & \big\langle m| \rho^k |2 \big\rangle & \ldots  & \big\langle m| \rho^k |m \big\rangle \\
\end{pmatrix},
\end{align}
where $m=d-1$ in which $d$ represents the dimension of the state and $k$ denotes the subsystem.
 
Let us consider certain non-Hermitian operators, $A_k^i$, $k=1,2$, $i=1,2,\ldots,$ $m,$ which act on $\rho^{k}$, are of the form \cite{hill2010}
\begin{align}  
\label{3} A_k^i ={|0\rangle}_k{\langle i|},~{A_{k}^{i}}^{\dagger} ={|i\rangle}_k{\langle 0|},~{A_k^{i} A_{k}^{i}}^{\dagger} ={|0\rangle}_k{\langle 0|},
~{A_{k}^{i}}^{\dagger} {A_k^{i}} = {|i\rangle}_k{\langle i|},~k=1,2. 
\end{align}
We observe that every element in the 
density matrix (\ref{3a}) can be expressed by the expectation value of 
the above non-Hermitian operators and their products. For example,  
\begin{align}
\begin{tabular}{ccc}
$\langle A_k^{i} {A_k^{i}}^{\dagger} \rangle = \langle 0|\rho^k| 0 \rangle$, & ~~~$\langle {A_k^{i}}^{\dagger} \rangle = \langle 0|\rho^k| i \rangle$,& ~~$\langle A_k^{i}\rangle = \langle i|\rho^k| 0 \rangle$,\\
$\langle {A_k^{i}}^{\dagger} A_k^{i} \rangle = \langle i|\rho^k| i \rangle$, & $\langle {A_k^{i}}^{\dagger} A_k^{1} \rangle = \langle 1|\rho^k| i \rangle$,
& $\langle {A_k^{i}}^{\dagger} A_k^{2} \rangle = \langle 2|\rho^k| i \rangle$
\end{tabular}
\end{align}  
and so on. In terms of these non-Hermitian operators ($A_k^i$'s) the matrix (\ref{3a}) reads 
 
\begin{align}
\label{3b} \rho_E^k= 
\begin{pmatrix}
\langle A_k^{1} {A_{k}^{1}}^{\dagger} \rangle  & \langle {A_{k}^{1}}^{\dagger} \rangle & \langle {A_{k}^{2}}^{\dagger} \rangle  &\ldots & \langle {A_{k}^{m}}^{\dagger} \rangle \\
\langle A_k^{1} \rangle  & \langle {A_{k}^{1}}^{\dagger}A_{k}^{1} \rangle & \langle {A_{k}^{2}}^{\dagger}A_{k}^{1} \rangle  &\ldots & \langle {A_{k}^{m}}^{\dagger}A_{k}^{1} \rangle \\
\langle A_k^{2} \rangle  & \langle {A_{k}^{1}}^{\dagger}A_k^{2} \rangle & \langle {A_{k}^{2}}^{\dagger}A_k^{2} \rangle  &\ldots & \langle {A_{k}^{m}}^{\dagger}A_k^{2} \rangle \\
\vdots & \vdots  & \vdots & \ddots & \vdots \\
\langle A_k^{m} \rangle  & \langle {A_{k}^{1}}^{\dagger}A_k^{m} \rangle & \langle {A_{k}^{2}}^{\dagger}A_k^{m} \rangle  & \ldots& \langle {A_{k}^{m}}^{\dagger}A_k^{m} \rangle \\
\end{pmatrix}. 
\end{align}

The matrix representation of a bipartite state in terms of the expectation value of 
the operators (\ref{3}) can be constructed by making the tensor product between expectation value matrices of first and second subsystems, that is
\begin{align}
\label{3c} \rho_E= 
\begin{pmatrix}
\langle A_1^{1} {A_{1}^{1}}^{\dagger} \rangle   & \ldots & \langle {A_{1}^{m_1}}^{\dagger} \rangle \\
\langle A_1^{1} \rangle  &  \ldots & \langle {A_{1}^{m_1}}^{\dagger}A_{1}^{1} \rangle \\
\vdots   & \ddots & \vdots \\
\langle A_1^{m_1} \rangle  &  \ldots& \langle {A_{1}^{m_1}}^{\dagger}A_1^{m_1} \rangle \\
\end{pmatrix}
\otimes
\begin{pmatrix}
\langle A_2^{1} {A_{2}^{1}}^{\dagger} \rangle & \ldots & \langle {A_{2}^{m_2}}^{\dagger} \rangle \\
\langle A_2^{1} \rangle  & \ldots & \langle {A_{2}^{m_2}}^{\dagger}A_{2}^{1} \rangle \\
\vdots  & \ddots  & \vdots \\
\langle A_2^{m_2} \rangle  &  \ldots& \langle {A_{2}^{m_2}}^{\dagger}A_2^{m_2} \rangle \\
\end{pmatrix}.
\end{align}
Expanding (\ref{3c}) we get 
\begin{align}
\label{3d}  \rho_E =
\begin{pmatrix}
\langle A_1^{1} {A_{1}^{1}}^{\dagger} A_2^{1} {A_{2}^{1}}^{\dagger} \rangle   
& \ldots & \langle A_1^{1} {A_{1}^{1}}^{\dagger}{A_{2}^{m_2}}^{\dagger} \rangle & \ldots & \langle {A_{1}^{m_1}}^{\dagger}{A_{2}^{m_2}}^{\dagger} \rangle  \\
\vdots  & \ddots & \vdots & \ddots & \vdots \\
\langle A_1^{1} {A_{1}^{1}}^{\dagger} A_2^{m_2} \rangle & \ldots & \langle A_1^{1} {A_{1}^{1}}^{\dagger}{A_{2}^{m_2}}^{\dagger}{A_{2}^{m_2}} \rangle 
& \ldots & \langle {A_{1}^{m_1}}^{\dagger}{A_{2}^{m_2}}^{\dagger}{A_{2}^{m_2}} \rangle \\
\langle A_1^{1} A_2^{1}{A_{2}^{1}}^{\dagger} \rangle & \ldots & \langle A_1^{1} {A_{2}^{m_2}}^{\dagger}\rangle & \ldots & \langle {A_{1}^{m_1}}^{\dagger}{A_{1}^{1}}{A_{2}^{m_2}}^{\dagger} \rangle \\
\vdots & \ddots & \vdots &  \ddots & \vdots \\
\langle A_1^{1} A_2^{m_2}\rangle & \ldots & \langle A_1^{1} {A_{2}^{m_2}}^{\dagger} A_2^{m_2} \rangle & \ldots & \langle {A_{1}^{m_1}}^{\dagger}{A_{1}^{1}}{A_{2}^{m_2}}^{\dagger}{A_{2}^{m_2}} \rangle \\
\vdots & \vdots & \vdots & \vdots & \vdots \\
\langle A_1^{m_1} A_2^{1}{A_{2}^{1}}^{\dagger} \rangle  & \ldots & \langle A_1^{m_1} {A_{2}^{m_2}}^{\dagger}\rangle & \ldots & \langle {A_{1}^{m_1}}^{\dagger}{A_{1}^{m_1}}{A_{2}^{m_2}}^{\dagger} \rangle \\
\vdots & \ddots & \vdots & \ddots & \vdots \\
\langle A_1^{m_1} A_2^{m_2}\rangle  & \ldots & \langle A_1^{m_1} {A_{2}^{m_2}}^{\dagger} A_2^{m_2} \rangle & \ldots & \langle {A_{1}^{m_1}}^{\dagger}{A_{1}^{m_1}}{A_{2}^{m_2}}^{\dagger}{A_{2}^{m_2}} \rangle
\end{pmatrix}.
\end{align}
Equation (\ref{3d}) is an equivalent representation of the density matrix of 
an $M\otimes N$ bipartite state, that is
\begin{align}
\label{rho} \rho= 
\begin{pmatrix}
\rho_{1,1} & \rho_{1,2} & \rho_{1,3} & \ldots & \rho_{1,n} \\
\rho_{2,1} & \rho_{2,2} & \rho_{2,3} & \ldots & \rho_{2,n} \\
\rho_{3,1} & \rho_{3,2} & \rho_{3,3} & \ldots & \rho_{3,n} \\
\vdots & \vdots  & \vdots  & \ddots & \vdots \\
\rho_{m,1} & \rho_{m,2} & \rho_{m,3} & \ldots & \rho_{m,n} 
\end{pmatrix}.
\end{align}

Comparing the matrix elements in (\ref{3d}) with (\ref{rho}) we observe that 
$\langle A_1^{1} {A_{1}^{1}}^{\dagger} A_2^{1}$ ${A_{2}^{1}}^{\dagger} 
\rangle $  yields  $\rho_{1,1}$, $\langle {A_{1}^{m_1}}^{\dagger}{A_{2}^{m_2}}^{\dagger} \rangle$  
yields  $\rho_{1,n}$ and so on. In other words, all the elements in the density matrix of 
bipartite states can now be represented by the expectation values of 
non-Hermitian operators $A_k^{i}$ and ${A_{k}^{i}}^{\dagger}$ and their products.
To illustrate this, let us consider a Bell state, $|\psi\rangle = a|00\rangle + b|11\rangle$ and its density operator 
$\rho = |a|^2|00\rangle \langle00| + ab^*|00\rangle \langle11| + a^*b |11\rangle \langle00| + |b|^2|11\rangle \langle11|$.
By using the ideas given above one can get  $|a|^2$ by computing $\langle A_1^1 {A_1^1}^{\dagger} A_2^1 {A_2^1}^{\dagger} \rangle $, 
$ab^*$ and $a^*b$  from the expectation values $\langle {A_1^1}^{\dagger} {A_2^1}^{\dagger} \rangle $ and 
$\langle {A_1^1} {A_2^1} \rangle $ respectively and $|b|^2$ by evaluating 
$\langle {A_1^1}^{\dagger}A_1^1 {A_2^1}^{\dagger}A_2^1 \rangle.$ 
Thus one can unambiguously represent   
every element in the density matrix by the expectation value of non-Hermitian operators and their products. 
In fact, with the aid of matrix (\ref{3d}) one can extract the value of any element of an arbitrary bipartite state. 
One can also construct this type of matrix for multipartite states. However, in this paper we confine our 
attention only on bipartite states.

We mention here that one can easily obtain the reduced density matrix from equation (\ref{3c})
instead of taking partial trace. It can be simply obtained by calculating all the expectation values present  
in any one of the susbsytems in (\ref{3c}), which is required. 

\section{\label{sec3} Utilization of expectation value matrix $\rho_E$ }
To demonstrate that any expression which involves matrix elements can be reformulated by expectation values of operators,  
we write Tr$(\rho^2)$ in terms of matrix elements for an arbitrary dimensional bipartite states. 
We then rewrite this expression in terms of expectatin values of 
operators. We reformulate the condition for pure product state from the 
expectation value of non-Hermitian operators 
into density matrix elements. In addition to the above, we present an operational 
form of partial transposition by employing the operators present in (\ref{3c}).
\subsection{Trace} 
Let us recall the condition $\mathrm{Tr}(\rho^2)=1$ for pure states,  
\begin{align}
\label{31} \textrm{Tr}(\rho^2) = \sum_{i=1}^{d_1\times d_2} \rho_{i,i}^2 + 2 \sum_{i=1}^{(d_1\times d_2)-1}~\sum_{j=i+1}^{d_1\times d_2}\rho_{i,j}~\rho_{j,i} = 1, 
\end{align}
where $d_1$ and $d_2$ represent the 
dimensions of the first and second subsystem respectively.
Equation (\ref{31}) can now be expressed solely in terms of the expectation values  
of non-Hermitian operators by comparing Eqs. (\ref{3d}) with (\ref{rho}). Replacing the elements by their expectation values 
of non-Hermitian operators $A_k^i$ and ${A_k^i}^{\dagger}$ and their products suitably we find 

\begin{align}
\text{Tr}(\rho^2) =& \langle A_1^1 {A_1^1}^{\dagger} A_2^1 {A_2^1}^{\dagger} \rangle^2 + \sum_{m_2=1}^{d_2-1}\langle A_1^1 {A_1^1}^{\dagger} 
{A_2^{m_2}}^{\dagger} A_2^{m_2}  \rangle^2  + \sum_{m_1=1}^{d_1-1}\langle {A_1^{m_1}}^{\dagger} A_1^{m_1} A_2^1 {A_2^1}^{\dagger}\rangle^2 \notag\\
& + \sum_{m_1=1}^{d_1-1} \sum_{m_2=1}^{d_2-1} \langle {A_1^{m_1}}^{\dagger} A_1^{m_1} {A_2^{m_2}}^{\dagger} A_2^{m_2}  \rangle^2 
+ 2 \Bigg\{ \sum_{m_2=1}^{d_2-1}  \langle {A_1^{1} A_1^{1}}^{\dagger} {A_2^{m_2}}^{\dagger} \rangle \langle {A_1^{1} A_1^{1}}^{\dagger} {A_2^{m_2}} \rangle \notag \\
& + \sum_{m_{2j}=1}^{d_2-2} \sum_{m_{2i}=m_{2j}+1}^{d_2-1} \langle A_1^{1} {A_1^{1}}^{\dagger} {A_2^{m_{2i}}}^{\dagger} A_2^{m_{2j}}\rangle  
\langle A_1^{1} {A_1^{1}}^{\dagger} {A_2^{m_{2j}}}^{\dagger} A_2^{m_{2i}}\rangle  \notag \\
&+ \sum_{m_1=1}^{d_1-1}\langle {A_1^{m_1}}^{\dagger} A_2^{1} {A_2^1}^{\dagger} \rangle \langle {A_1^{m_1}} A_2^{1}{A_2^1}^{\dagger} \rangle 
+ \sum_{m_1=1}^{d_1-1}\sum_{m_2=1}^{d_2-1} \langle {A_1^{m_1}}^{\dagger} A_1^{m_1} {A_2^{m_2}}^{\dagger} \rangle \langle {A_1^{m_1}}^{\dagger} A_1^{m_1} {A_2^{m_2}} \rangle \notag 
\end{align}
\begin{align}
& + \sum_{m_{1j}=1}^{d_1-2} \sum_{m_{1i}=m_{1j}+1}^{d_1-1} \langle {A_1^{m_{1i}}}^{\dagger} A_1^{m_{1j}}A_2^{1} {A_2^{1}}^{\dagger} \rangle 
\langle {A_1^{m_{1j}}}^{\dagger} A_1^{m_{1i}}A_2^{1} {A_2^{1}}^{\dagger} \rangle \notag \\
& +  \sum_{m_1=1}^{d_1-1} \sum_{m_{2j}=1}^{d_2-2} \sum_{m_{2i}=m_{2j}+1}^{d_2-1} \langle {A_1^{m_1}}^{\dagger} A_1^{m_1} {A_2^{m_{2i}}}^{\dagger} 
A_2^{m_{2j}}\rangle \langle {A_1^{m_1}}^{\dagger} A_1^{m_1} {A_2^{m_{2j}}}^{\dagger} A_2^{m_{2i}}\rangle \notag \\
& + \sum_{m_2=1}^{d_2-1} \sum_{m_1=1}^{d_1-1} \langle {A_1^{m_1}}^{\dagger} {A_2^{m_2}}^{\dagger}A_2^{m_2} \rangle \langle {A_1^{m_1}} {A_2^{m_2}}^{\dagger}A_2^{m_2} \rangle\notag \\
& +  \sum_{m_2=1}^{d_2-1} \sum_{m_{1j}=1}^{d_1-2} \sum_{m_{1i}=m_{1j}+1}^{d_1-1} \langle {A_1^{m_{1i}}}^{\dagger} A_1^{m_{1j}} {A_2^{m_2}}^{\dagger} 
A_2^{m_2}\rangle \langle {A_1^{m_{1j}}}^{\dagger} A_1^{m_{1i}} {A_2^{m_2}}^{\dagger} A_2^{m_2}\rangle \notag \\
&+ \sum_{m_1=1}^{d_1-1} \sum_{m_2=1}^{d_2-1} \langle {A_1^{m_1}}^{\dagger} {A_2^{m_2}}^{\dagger} \rangle  \langle {A_1^{m_1}} {A_2^{m_2}} \rangle 
 + \sum_{m_1=1}^{d_1-1} \sum_{m_2=1}^{d_2-1} \langle {A_1^{m_1}}^{\dagger} {A_2^{m_2}} \rangle \langle {A_1^{m_1}} {A_2^{m_2}}^{\dagger} \rangle 
\notag \\
&+ \sum_{m_1=1}^{d_1-1} \sum_{m_{2j}=1}^{d_2-2} \sum_{m_{2i}=m_{2j}+1}^{d_2-1} \langle {A_1^{m_1}}^{\dagger} {A_2^{m_{2i}}}^{\dagger} 
A_2^{m_{2j}}\rangle \langle {A_1^{m_1}} {A_2^{m_{2j}}}^{\dagger} A_2^{m_{2i}}\rangle  \notag \\
& + \sum_{m_1=1}^{d_1-1} \sum_{m_{2j}=1}^{d_2-2} \sum_{m_{2i}=m_{2j}+1}^{d_2-1} \langle {A_1^{m_1}}^{\dagger} {A_2^{m_{2j}}}^{\dagger} 
A_2^{m_{2i}}\rangle \langle {A_1^{m_1}} {A_2^{m_{2i}}}^{\dagger} A_2^{m_{2j}}\rangle \notag \\
& + \sum_{m_2=1}^{d_2-1} \sum_{m_{1j}=1}^{d_1-2} \sum_{m_{1i}=m_{1j}+1}^{d_1-1} \langle {A_1^{m_{1i}}}^{\dagger} A_1^{m_{1j}} {A_2^{m_2}}^{\dagger} \rangle \langle {A_1^{m_{1j}}}^{\dagger} A_1^{m_{1i}} {A_2^{m_2}} \rangle \notag \\
& + \sum_{m_2=1}^{d_2-1} \sum_{m_{1j}=1}^{d_1-2} \sum_{m_{1i}=m_{1j}+1}^{d_1-1} \langle {A_1^{m_{1i}}}^{\dagger} A_1^{m_{1j}} {A_2^{m_2}} \rangle 
\langle {A_1^{m_{1j}}}^{\dagger} A_1^{m_{1i}} {A_2^{m_2}}^{\dagger} \rangle \notag \\
& +  \sum_{m_{1i}=1}^{d_1-2} \sum_{m_{1j}=m_{1i}+1}^{d_1-1} \sum_{m_{2i}=1}^{d_2-2} \sum_{m_{2j}=m_{2i}+1}^{d_2-1}  \langle {A_1^{m_{1j}}}^{\dagger} 
A_1^{m_{1i}} {A_2^{m_{2j}}}^{\dagger} A_2^{m_{2i}} \rangle  \langle {A_1^{m_{1i}}}^{\dagger} A_1^{m_{1j}} {A_2^{m_{2i}}}^{\dagger} A_2^{m_{2j}} \rangle \notag \\
& + \sum_{m_{1i}=1}^{d_1-2} \sum_{m_{1j}=m_{1i}+1}^{d_1-1} \sum_{m_{2i}=1}^{d_2-2} \sum_{m_{2j}=m_{2i}+1}^{d_2-1}  \langle {A_1^{m_{1j}}}^{\dagger} 
A_1^{m_{1i}} {A_2^{m_{2i}}}^{\dagger} A_2^{m_{2j}} \rangle  \langle {A_1^{m_{1i}}}^{\dagger} A_1^{m_{1j}} {A_2^{m_{2j}}}^{\dagger} A_2^{m_{2i}} \rangle\Bigg\}.
\end{align}
The above expression is the general form of Tr$(\rho^2)$ for an arbitrary bipartite states. 
\subsection{Pure separable state}
In the case of pure separable states the expectation value of a joint measurement of 
two operators should be equal to the product of expectation value of individual 
measurement of the same operators acting on the  bipartite state, that is 
$\langle AB \rangle = \langle A\rangle \langle B\rangle$. 
Considering the operators that appear in the diagonal in (\ref{3d}) and imposing 
the pure separability condition given above and comparing it 
with (\ref{rho}), we can express the criterion of pure separable states in terms of density matrix elements. 
In the case of $M \otimes N$ pure product state the condition reads  
\begin{align}
\label{rhokk}\rho_{k,k} = \left(\sum_{i=(a\times d_2)+1}^{(a+1)\times d_2} \rho_{i,i} \right) \times \left( \sum_{i\in B} \rho_{i,i} \right),
\end{align}
where $(a\times d_2)+1 \leq k \leq (a+1) \times d_2$, $a=0,1,2,...,d_1-1$, 
~$B=\{i = (j\times d_2) +k ~|~ j=0,1,2,...,d_1-1 \}$ and $i=i-(d_1 \times d_2)$ iff $i>(d_1\times d_2)$. 
The value of $a$ can be determined by simply fixing the $a$ value for $k$ in 
the interval $(a\times d_2)+1 \leq k \leq (a+1) \times d_2$.
With the known value of $a$ one can proceed to check the separability in 
pure arbitrary bipartite states. 

If the given state does not satisfy the condition (\ref{rhokk}) then it 
should be a pure entangled one.
\subsection{Partial Transposition operation}
In the following, we point out how we can reformulate the partial transposition in terms of non-Hermitian 
operators and their products. To begin with, we consider a two qubit case and express 
the partial transposition operation. In this case, we find 
\begin{align}
\label{rph2q1} \rho_{2\otimes 2}^{T_1} =& A_1^1 {A_1^1}^{\dagger} \rho A_1^1 {A_1^1}^{\dagger} + A_1^1 \rho A_1^1 + {A_1^1}^{\dagger} \rho {A_1^1}^{\dagger}+ 
{A_1^1}^{\dagger}A_1^1  \rho {A_1^1}^{\dagger}A_1^1, \\
\label{rph2q2} \rho_{2\otimes 2}^{T_2} =& A_2^1 {A_2^1}^{\dagger} \rho A_2^1 {A_2^1}^{\dagger} + A_2^1 \rho A_2^1 + {A_2^1}^{\dagger} \rho {A_2^1}^{\dagger}+ 
{A_2^1}^{\dagger}A_2^1  \rho {A_2^1}^{\dagger}A_2^1. 
\end{align}
We can rewrite the above two expressions as  
\begin{align}
\label{rph2qk} \rho_{2\otimes 2}^{T_k} = A_k^1 {A_k^1}^{\dagger} \rho A_k^1 {A_k^1}^{\dagger} + A_k^1 \rho A_k^1 + {A_k^1}^{\dagger} \rho {A_k^1}^{\dagger}+ 
{A_k^1}^{\dagger}A_k^1  \rho {A_k^1}^{\dagger}A_k^1, 
\end{align}
where $k=1$ and $2$ represent the partial transposition with respect to first and second 
subsystem respectively. We note here that the operators appearing in the expressions 
(\ref{rph2q1}) and (\ref{rph2q2}) are taken from the expectation value matrix 
$\rho_E$ (\ref{3c}) (restricted to $2\otimes 2$). In other words, we have considered 
all the operators in the first and second subsystems.  

Equation (\ref{rph2qk}) can be generalized to an arbitrary bipartite state by considering 
all the operators present in the corresponding expectation value 
matrix (\ref{3c}), that is ($k=1$ and $2$ correspond to the transposition)  
\begin{align}
 \rho_{M\otimes N}^{T_k} = &A_k^1 {A_k^1}^{\dagger} \rho A_k^1 {A_k^1}^{\dagger} + {A_k^1}^{\dagger}  \rho {A_k^1}^{\dagger} + {A_k^2}^{\dagger}  \rho {A_k^2}^{\dagger} \notag \\
\label{ptd} &  + \cdots + {A_k^{{m_k}-1}}^{\dagger} A_k^{m_k} \rho  {A_k^{{m_k}-1}}^{\dagger} A_k^{m_k} +  {A_k^{m_k}}^{\dagger} A_k^{m_k} \rho  
{A_k^{m_k}}^{\dagger} A_k^{m_k},   
\end{align}
where $m_k=d_k-1$ corresponds to the subsystem $k$. Eq.(\ref{ptd}) is an operational form of partial transposition. 
If the eigenvalues of the density matrix $\rho_{M\otimes N}^{T_k}$ are positive then the 
underlying state is separable in view of PPT criterion \cite{peres1996}.
\section{Separability condition}
\label{sec4}
In Refs.\cite{guhne2010,gao2011} the biseparability and full separability criteria for $n$-partite quantum states using 
elements of density matrices were derived. 
In this work, we derive a separability criterion applicable to $C^d\otimes C^d$ bipartite states 
using the ideas given in the references \cite{guhne2010,gao2011}. 
To begin with, we consider a pure two qubit separable state 
$|\psi\rangle = (a |0\rangle + b |1\rangle) \otimes (c |0\rangle + d |1\rangle)$. 
For this state, we have $|\rho_{1,4}|= \sqrt{\rho_{1,1} \rho_{4,4}}$ or 
$|\rho_{1,4}|= \sqrt{\rho_{2,2} \rho_{3,3}}$ and $|\rho_{2,3}|= \sqrt{\rho_{1,1} \rho_{4,4}}$ 
or $|\rho_{2,3}|= \sqrt{\rho_{2,2} \rho_{3,3}}$. The first two expressions yield   
\begin{align}
2 |\rho_{1,4}| = \sqrt{\rho_{1,1} \rho_{4,4}} + \sqrt{\rho_{2,2} \rho_{3,3}}.  
\end{align}
Using the inequality of arithmetic and geometric mean, the above expression can be rewritten as   
\begin{align}
\label{rho14} 4 |\rho_{1,4}| \leq& (\rho_{1,1}+\rho_{4,4}) + 2 \sqrt{\rho_{2,2} \rho_{3,3}}. 
\end{align}
Similarly, for the element $|\rho_{2,3}|$, we find 
\begin{align}
\label{rho23} 4 |\rho_{2,3}| \leq (\rho_{1,1}+\rho_{4,4}) + 2 \sqrt{\rho_{2,2} \rho_{3,3}}.
\end{align}
From these two expressions, (\ref{rho14}) and (\ref{rho23}), we can extract the following condition 
\begin{align}
\label{1423} 4 \max \left\{|\rho_{1,4}|,|\rho_{2,3}| \right\} \leq (\rho_{1,1}+\rho_{4,4}) + 2 \sqrt{\rho_{2,2} \rho_{3,3}}.
\end{align}
One may also come across this type of inequalities in the partially separable multiqubit states \cite{seevinck2008}.

Now we generalize the  above condition to ${C}^d \otimes {C}^d$ bipartite states. 
In the two qubit case, we have derived the separability condition using the elements $|\rho_{1,4}|$ and $|\rho_{2,3}|$.  
In the higher dimensional case, we have a difficulty with which elements are to be measured. 
To choose the relevant elements,  we consider expectation value of joint measurement opertors $A^{\dagger} B^{\dagger} / A B$  
and its partially transposed and conjugated operator $ A^{\dagger} B / A B^{\dagger} $. 
We note here that to extract the required elements, we consider only $\langle A^{\dagger} B^{\dagger} \rangle$ and $\langle A^{\dagger} B \rangle$ 
which in turn provide non-zero off diagonal elements in Werner and isotropic classes of states \cite{chrus2006}.

Before demonstrating how to extract the required elements by substituting the non-Hermitian operators 
$A$ and $B$, we justify the proposed expectation values, that is $\langle A^{\dagger} B^{\dagger} \rangle$ and $\langle A^{\dagger} B \rangle$,  
by considering two qubit states. 
Substituting the non-Hermitian operator $A=|0\rangle\langle 1|$, with $A=B$ and comparing the matrix (\ref{3d}) with (\ref{rho}), 
we can prove that the operators   
$\langle A^{\dagger} B^{\dagger} \rangle$ and $\langle A^{\dagger} B \rangle$ give the matrix elements $|\rho_{1,4}|$ 
and $|\rho_{2,3}|$ respectively. One may note that only these two elements appear in the entanglement condition (\ref{1423}). 
With this verification now we proceed to construct separability condition for the higher dimensional states.

To begin with, we constitute the separability condition for the two qutrit state. We then generalize it to the two qudit states.  
In the two qutrit $(|\psi\rangle = (a |0\rangle + b |1\rangle + c |2\rangle) \otimes (d |0\rangle + e |1\rangle + f |2\rangle))$ separable case, 
we consider non-Hermitian operators $A$ and $B$ are of the form $|0\rangle \langle 1|, |0\rangle \langle 2|,$ and $|1\rangle 
\langle 2|$ which are the possible operators in $|0\rangle, |1\rangle, |2\rangle$ bases. Considering $A=B$ and 
substituting $|0\rangle \langle 1|, |0\rangle \langle 2|$ and 
$|1\rangle \langle 2|$ in $\langle A^{\dagger} B^{\dagger} \rangle$ and comparing the matrix (\ref{3d}) with (\ref{rho}), 
we find that $\langle A^{\dagger} B^{\dagger} \rangle$ produces the matrix elements 
$|\rho_{1,5}|$, $|\rho_{1,9}|$ and $|\rho_{5,9}|$. In the present case, the underlying inequalities read  
\begin{align}
2 |\rho_{1,5}| \leq \frac{\rho_{1,1}+\rho_{5,5}}{2} + \sqrt{\rho_{2,2} \rho_{4,4}}, \notag \\ 
2 |\rho_{1,9}| \leq \frac{\rho_{1,1}+\rho_{9,9}}{2} + \sqrt{\rho_{3,3} \rho_{7,7}},  \notag  \\ 
2 |\rho_{5,9}| \leq \frac{\rho_{5,5}+\rho_{9,9}}{2} + \sqrt{\rho_{6,6} \rho_{8,8}}.   
\end{align}
Adding and simplifying the above expressions, we get  
\begin{align}
\label{tqt1} 2 \left( |\rho_{1,5}|+|\rho_{1,9}|+|\rho_{5,9}| \right) \leq &~\rho_{1,1}+\rho_{5,5}+\rho_{9,9} \notag \\ 
&~+ \left(\sqrt{\rho_{2,2} \rho_{4,4}} + \sqrt{\rho_{3,3} \rho_{7,7}} + \sqrt{\rho_{6,6} \rho_{8,8}} \right).
\end{align}
Repeating the above analysis for the case $\langle A^{\dagger} B \rangle$ we end up with the following inequality, namely  
\begin{align}
 \label{tqt2}2 \left( |\rho_{2,4}|+|\rho_{3,7}|+|\rho_{6,8}| \right) \leq &~\rho_{1,1}+\rho_{5,5}+\rho_{9,9} \notag \\
&~+ \left(\sqrt{\rho_{2,2} \rho_{4,4}} + \sqrt{\rho_{3,3} \rho_{7,7}} + \sqrt{\rho_{6,6} \rho_{8,8}} \right). 
\end{align}
Combining the equations (\ref{tqt1}) and (\ref{tqt2}) suitably we obtain the separability condition for the two qutrit states in 
the form
\begin{align}
2 \max &\left\{\left( |\rho_{1,5}|+|\rho_{1,9}|+|\rho_{5,9}| \right),\left( |\rho_{2,4}|+|\rho_{3,7}|+|\rho_{6,8}| \right)\right\} \notag \\
& \qquad \qquad \qquad \qquad \qquad  \leq (\rho_{1,1}+\rho_{5,5}+\rho_{9,9}) +  \big(\sqrt{\rho_{2,2} \rho_{4,4}} \notag \\
&  \qquad \qquad \qquad \qquad \qquad \quad + \sqrt{\rho_{3,3} \rho_{7,7}} + \sqrt{\rho_{6,6} \rho_{8,8}} \big).
\end{align}

Now we generalize the above conditions to the two qudit states. Substituting the non-Hermitian operators 
$|a_1\rangle\langle a_2|$, $|a_1\rangle\langle a_3|$, \ldots, $|a_1\rangle\langle a_n|$, $|a_2\rangle\langle a_3|$, \ldots, 
$|a_2\rangle\langle a_n|$, \ldots, and $|a_{n-1}\rangle\langle a_n|$, which are in the 
$|a_1\rangle$, $|a_2\rangle$, $|a_3\rangle$, \ldots, $|a_n\rangle$ bases, in the above expectation values of operators and following 
the procedure given above, we arrive at  
\begin{align}
\label{cond} 2 \times &\max \left\{ \sum_{0\leq i < j \leq (d-1)} |\rho_{i(d+1)+1,~j(d+1)+1}|, \sum_{0\leq i < j \leq (d-1)} |\rho_{id+j+1,~jd+i+1}| \right\}  \notag \\ 
&  \qquad \qquad \quad \leq\frac{(d-1)}{2} \sum_{i=0}^{d-1} \rho_{i(d+1)+1, i(d+1)+1} \notag \\
&  \qquad \qquad \quad \quad+ \left( \sum_{0\leq i < j \leq (d-1)} \sqrt{\rho_{id+j+1,~id+j+1} \times \rho_{jd+i+1,~jd+i+1} } \right),  
\end{align}
where $\rho_{p,q}$ represents the density matrix element in $p^{\text{th}}$ row and $q^{\text{th}}$ column and $d$ represents the dimension of a qudit state 
in ${C}^d \otimes {C}^d$ bipartite states. Using some simple algebras and Cauchy inequality we can extend the condition (\ref{cond}) 
to mixed bipartite states as given in Ref.\cite{gao2011}. 
The inequality (\ref{cond}) holds for mixed separable states  and violation of this inequality implies 
that the state is  entangled. As we have seen in Sect.\ref{sec3}, any expression involves density matrix elements can be 
reformulated by the expectation values given in (\ref{3d}). 
Therefore, the experimental accessibility of the criterion (\ref{cond}) also becomes possible \cite{toth2003,guhne2010,thew2002,pati2014}.  
\section{Examples}
\label{sec5}
In this section, we illustrate the above ideas by considering various mixed states. \\

\noindent 1. We consider a $d \otimes d$ Werner state parametrized by $\eta \in \mathbb{R}$ \cite{werner1989,jones2007}.
\begin{equation}
W_d^{\eta} = \left(\frac{d-1+\eta}{d-1}\right) \frac{I}{d^2} - \left(\frac{\eta}{d-1} \right) \frac{V}{d}, \label{wer}
\end{equation}
where $I$ is the identity operator, $ V=\sum_{i,j}^d |ij\rangle \langle ji|$ is the flip operator, $d$ is the dimension and $0\leq \eta\leq 1$. 
For $\eta = 0$, it becomes a separable state.  
Werner states are entangled for $\eta >1/(d+1)$ \cite{werner1989,jones2007}.
To identify the entangled regions of the above $d\otimes d$ Werner states for various $d$ values, 
we derive the following general function, 
\begin{align}
p =&  \bigg[\frac{d-1+\eta-\eta d}{2 d} +\left(\sum_{i=1}^d (d-i)\right) \times \left(\frac{d-1+\eta}{(d-1)d^2}\right)\bigg] \notag\\
&\times \left(\frac{1}{2\left(\sum_{i=1}^d (d-i)\right)\times \frac{\eta}{(d-1)d}}\right),   
\end{align}
by applying the inequality (\ref{cond}) on (\ref{wer}). We plot the outcome in fig.1(a). 
For a given dimension, the value of this function is less than one, that is $p<1$, 
then the state is entangled. We also employ this function to study the higher dimensional Werner states (not included in figure). 
Our result shows that the entangled region keep on decreasing upto dimension $d =134$ and 
above this dimension, the entangled region becomes constant which in turn lies between $\eta=0.665$ and $\eta=1$. \\

\begin{figure}[ht]
\begin{center}
\includegraphics[width=0.85\textwidth]{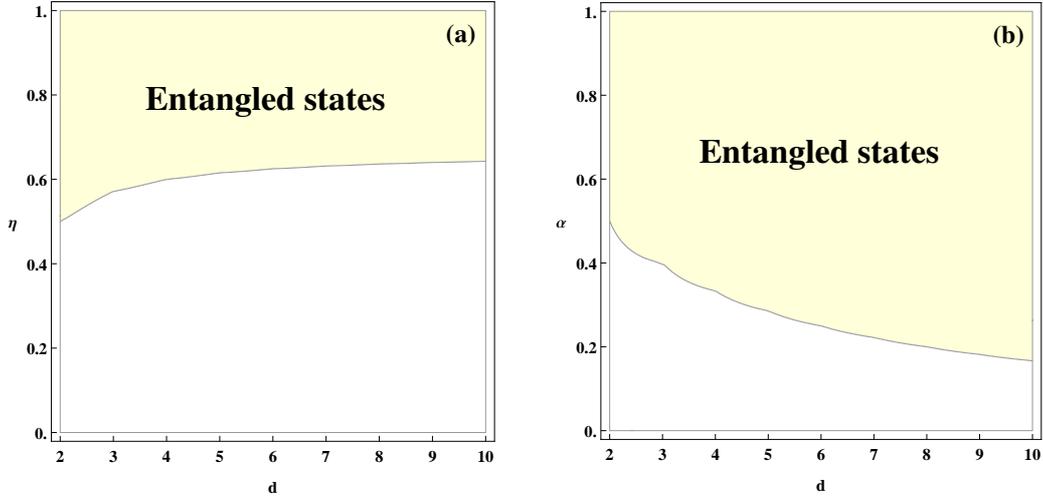}
\caption{Entanglement region of $d\otimes d$ (a) Werner states and (b) Isotropic states} 
\end{center}
\end{figure}

\noindent 2. The isotropic states, which were introduced in Ref. \cite{horo1999}, can be written as the mixtures of the maximally mixed state and 
maximally entangled state $|\psi_{+}\rangle = \frac{1}{\sqrt{d}} \sum_{i=0}^{d-1} |ii\rangle$, that is 
\begin{equation}
\label{rhalp}\rho_{\alpha} = \frac{(1-\alpha)}{d^2} I + \alpha |\psi_+\rangle \langle \psi_+|, 
\end{equation}
where $0\leq \alpha \leq 1$ and $d$ is the dimension of states. Here also we aim to identify the entangled region of isotropic states. 
As we did in the previous case, we construct a function 
\begin{align} 
q = \bigg[\frac{(d-1)d}{2}\times\left(\frac{1-\alpha}{d^2}+\frac{\alpha}{d}\right) + \left(\sum_{i=1}^{d}(d-i)\right)\times\frac{1-\alpha}{d^2}\bigg]\times \frac{1}{2\times\left(\displaystyle\sum_{i=1}^{d}(d-i)\right)\frac{\alpha}{d}}
\end{align}
from the inequality (\ref{cond}) for the $d\otimes d$ isotropic states (\ref{rhalp}) and depict the result in fig.1(b).  
If $q<1$, then the state is entangled. 
We have also evaluated the function $q$ for higher dimensions $(d >1000)$ and observed that the 
entangled region of (\ref{rhalp}) keep on increasing but the states are not entangled at $\alpha=0$. 
\\ \\ 
3. We consider another state \cite{li2012}, which is a mixture of PPT entangled state and maximal entangled state, namely 
\begin{equation}
\label{pptme}\rho_p = (1-p) \rho_a + p P_+ ,
\end{equation}
where 
\begin{eqnarray}
\label{pptex}\rho_a = \frac{8a}{8a+1} \rho_{\mbox{insep}} + \frac{1}{8a+1} P_{\Phi_a}, 
\end{eqnarray}
$\rho_{\mbox{insep}} = \frac{3}{8} P_{\Psi} + \frac{1}{8} Q$, $\Phi_a = e_3 \otimes \left( \sqrt{\frac{1+a}{2}} e_1 + \sqrt{\frac{1-a}{2}} e_3 \right)$,  $Q = I \otimes I - \big( \sum_{i=1}^3 P_{e_i}$ $\otimes P_{e_i} + P_{e_3} \otimes P_{e_1} \big)$,   $\Psi = \frac{1}{\sqrt{3}} (e_1 \otimes e_1 + e_2 \otimes e_2 + e_3 \otimes e_3)$,  $0\leq a\leq 1 $ and $P_{+} = |\psi_+\rangle \langle\psi_+|$, $|\psi_+\rangle = \frac{1}{\sqrt{3}} \sum_{i=0}^2 |ii\rangle$.  The separability criterion in terms of the range of the density matrix detects the entanglement in (\ref{pptex}) for $a\neq 0,1$ \cite{horo1997}, in which $\rho_{\mbox{insep}}$ violates the condition (\ref{cond}), indicating that the state is inseparable. 
The inequalities proposed in Ref.\cite{li2012} detects the entanglement for the whole region of $0<p\leq 1$ at  $a=0.236$ for (\ref{pptme}). 
Our inequality (\ref{cond}) detects the entanglement for $0<a<1$ and $0<p\leq 1$.  
It also shows that while the parameter value $a$ increases, minimum value of $p$, to be entangled, is decreasing. 
\\ \\ 
4. Finally, we consider a state which is a mixture of $3\times 3$ state (form a Unextendible Product Bases) and the maximal entangled singlet \cite{li2012}, that is 
\begin{equation}
\label{upbp}\rho_p = (1-p) \rho + p P_+, 
\end{equation}
where ${\displaystyle \rho = \frac{1}{4} \left( I - \sum_{i=0}^4 |\xi_i\rangle\langle \xi_i|\right)}$,  $|\xi_0\rangle = \frac{1}{\sqrt{2}} |0\rangle (|0\rangle - |1\rangle)$, $|\xi_1\rangle = \frac{1}{\sqrt{2}} (|0\rangle - |1\rangle) |2\rangle$, $|\xi_2\rangle = \frac{1}{\sqrt{2}} |2\rangle (|1\rangle - |2\rangle)$, $|\xi_3\rangle = \frac{1}{\sqrt{2}} (|1\rangle - |2\rangle) |0\rangle$ and $|\xi_4\rangle = \frac{1}{3} (|0\rangle +|1\rangle +|2\rangle)(|0\rangle +|1\rangle +|2\rangle)$ \cite{bennett1999}, which is entangled according to the realignment criterion \cite{chen2003}. For the state (\ref{upbp}), the Bell inequality \cite{li2010} detects the entanglement for $0.57602 \leq p \leq 1$ and inequality given in Ref.\cite{li2012} detects the entanglement for $0.18221 \leq p \leq 1$. Our condition (\ref{cond}) detects the entanglement for $0.44 \leq p \leq 1$. 
\section{\label{sec6}Conclusion}
In this paper, we have exploited the utility of non-Hermitian operators to characterize the bipartite states. 
We have constructed a new form of density matrix whose elements are expressed in terms of expectation values of non-Hermitian operators and their products from which we have shown that the condition which involves matrix elements can be reformulated in terms of  expectation values of operators and vice versa. 
We then derived the separability condition for ${C}^d \otimes {C}^d$ bipartite states in terms of density matrix elements using non-Hermitian operators and it can be reformulated in the form of expectation values of non-Hermitian operators and their products. 
We have utilized our condition to detect entanglement in $d\otimes d$ Werner states, $d\otimes d$ Isotropic states, 
$3\otimes 3$ PPT entangled state and Unextendible product bases mixed with maximally entangled state. 
Through this work we have brought out the utility of non-Hermitian operators in identifying each and every element of the given state and demonstrated 
how they are useful in detecting entanglement. The application of this procedure to multipartite states is under progress.

\end{document}